\documentclass{pramana}
\usepackage[T1]{fontenc}
\usepackage{graphicx}
\usepackage{amsmath,bm}
\usepackage{amssymb}
\usepackage{xcolor}
\usepackage{hyperref}
\begin{document}

\title{Viscous damping of r-modes and emission of gravitational waves}

\author{Debasis Atta$^{1}$ and D. N. Basu$^{2}$\textsuperscript{*}}

\affilOne{$^{1}$Department of Higher Education, Government of West Bengal, Bikash Bhavan, Saltlake, Kolkata 700091, India \\}

\affilTwo{$^{2}$Variable  Energy  Cyclotron Centre, 1/AF Bidhan Nagar, Kolkata 700064, India}

\twocolumn[{

\maketitle

\corres{dnb@vecc.gov.in}

\msinfo{4 October 2024}{8 January 2025}{22 January 2025}

\begin{abstract}

    The Rossby mode (r-mode) perturbation in pulsars as a steady gravitational wave (GW) source has been explored. The effect of a rigid crust on viscous damping and dissipation rate in the boundary layer between fluid core and crust has been studied. The intensity of the emitted GWs in terms of the strain tensor amplitude has been estimated with the approximation of slow rotation using equation of state derived from the APR and Skyrme effective interactions with Brussels-Montreal parameter sets. The core of the neutron star has been considered to be $\beta$-equilibrated nuclear matter containing neutrons, protons, electrons and muons, surrounded by a solid crust. Calculations have been made for critical frequencies, the time evolution and the rate of change of the spin frequencies across a broad spectrum of pulsar masses.

\end{abstract}


}]

\maketitle

\noindent
\section{Introduction}
\label{Section 1} 
     
    Following the groundbreaking research conducted by Baade–Zwicky \cite{Ba34} and Oppenheimer–Volkoff \cite{Op39}, alongside the discovery of pulsars as rotating neutron stars (NSs) nearly sixty years ago \cite{Go68}, numerous investigations were carried out to unravel the mysteries surrounding the structure of NSs. It is now widely acknowledged that the density variation from the surface to the core spans approximately fifteen orders of magnitude. In recent years quite a few NSs with mass $M>2M_\odot$ have been observed \cite{Bassa2017,Miller2021}. Their rotations generate quasi-normal modes that can act as probes for the effects of general relativity such as gravitational waves (GWs) and the properties of ultradense matter. 
    
    A pulsar's density domain can be broadly divided into two regions: a crust that makes up about 0.5$\%$ of the star's mass and nearly 10$\%$ of its radius. The remainder of the star's mass and radius are accounted for by the core. The domain between the neutron drip point and the inner edge that divides the solid crust from the uniform liquid core is known as the inner crust. A phase transition from the high-density homogeneous matter to the inhomogeneous one at lower densities takes place at the inner edge. When the uniform neutron-proton-electron-muon ($npe\mu$) matter becomes unstable with regard to the separation into two coexisting phases (one corresponding to nuclei, the other to a nucleonic sea), the transition density reaches its critical value \cite{La07}. The core-crust transition takes place in the inner crust, which is made up of a crystal lattice of nuclei submerged in a neutron superfluid. The superfluid containing neutrons (both inside the inner crust and deeper inside the star) is entangled with a regular pattern of spinning vortices produced by the rotation of a pulsar.
    
    The Rossby mode (r-mode) non-radial perturbation is a rotation-powered oscillation mode where the Coriolis force due to the star's rotation acts as a restoring force. Its unique feature is that the flow pattern is in the opposite direction of the pulsar's rotation as seen by an observer at rest on the star (retrograde) and in the same direction as the star's rotation when observed by an observer at infinity (prograde). Since the emitted GWs from non-radial disturbances carries energy, it should lead to the damping of the perturbation. However, Chandrasekhar, Friedman and Schutz demonstrated \cite{Ch70,Fr78} that GW radiation emitted as a result of any perturbation that is retrograde in the co-rotating frame and progarde in the inertial frame would accelerate rather than lessen the perturbation. The gravitational wave emission from r-mode oscillations in pulsars is suppressed by various viscous damping, which may remain effective up to 10$^4$ years after mass accretion by the star stops. The instability of the r-mode is significant only if its growth rate exceeds the rate at which it is dampened by viscosity. Thus, the time scale for gravitationally driven instabilities must be significantly shorter than the time scale for viscous damping.   
    
    The variations in the period of rotation of compact stars with time can reveal their internal fluctuations such as the core-crust coupling and decoupling. The GWs from pulsars due to their rotational instabilities can explore the high density behavior of an Equation of State (EoS). Indeed, our knowledge of compact stars is largely dependent on data obtained from GW emissions observed by LIGO and Virgo collaborations \cite{Miller2019,Abbott2019}. Significant progress has been made in understanding the properties of compact stars, such as their mass limits, angular momenta, creation rates and spin evolution mechanism. 
    
    The instabilities in rotation of compact stars can directly be correlated to the unstable modes \cite{Andersson1998,Andersson2003,Freidman1998,Provost1981,Andersson2001,Bondarescu2007} of oscillations. In the present work, the r-mode instability has been studied in the context of pulsars using EoSs obtained from the effective nucleon-nucleon (NN) interaction described by APR \cite{Ak98} and Brussels-Montreal Skyrme effective interaction with the BSk22, BSk24 and BSk26 parameter sets \cite{Gor13}. The time evolution of spin-frequency and the spin-down rate along with the variation of critical frequencies with respect to temperature have been calculated across a broad spectrum of pulsar masses. With the approximation of slow rotation, the intensity of the emitted GWs in terms of the strain tensor amplitude has also been estimated.

\begin{figure}[t]
\vspace{-0.0cm}
\includegraphics[width=8cm]{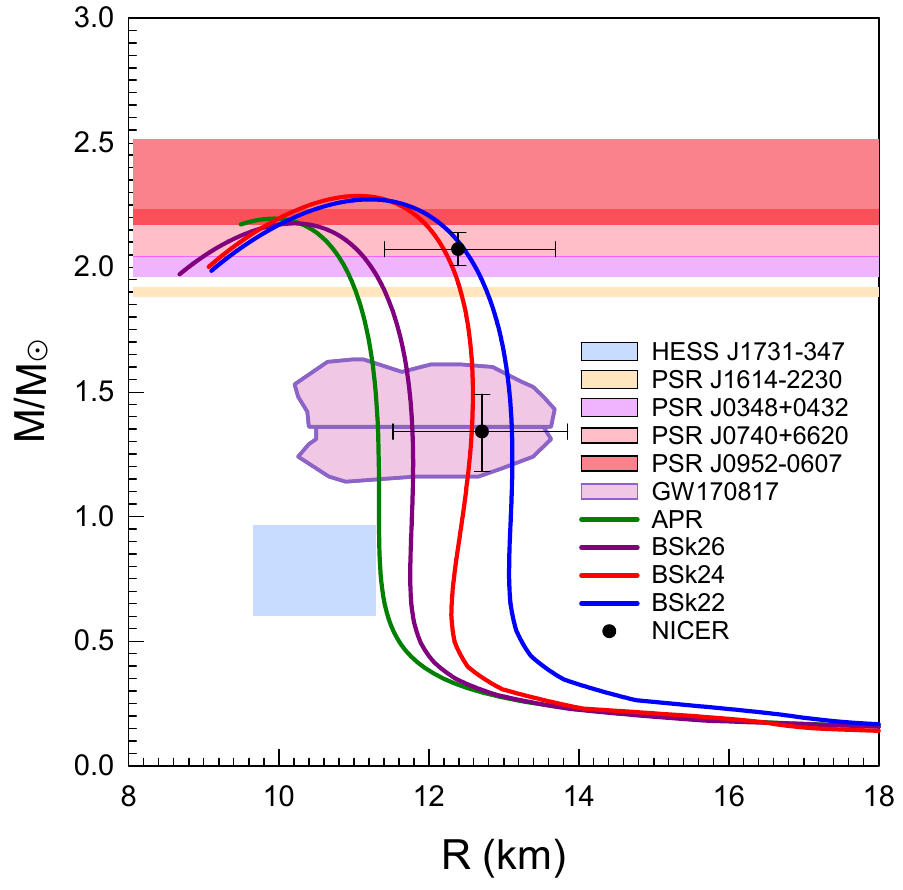}
\caption{The mass-radius plots for APR, BSk22, BSk24 and BSk26 EoSs. The shaded regions represent the HESS J1731-347 remnant \cite{Dor22}, the GW170817 event \cite{Abbott19}, PSR J1614-2230 \cite{Arz18}, PSR J0348+0432 \cite{Ant13}, PSR J0740+6620 \cite{Cro20}, and PSR J0952-0607 \cite{Rom22} pulsar observations for the possible maximum mass. The black dots with error bars represent recent radius measurements from NICER \cite{Ril21,Ril19}.}
\label{fig1}
\vspace{-0.0cm}
\end{figure}

\noindent
\section{Non-radial perturbations in rotating neutron stars} 
\label{Section 2}
 
    The r-mode oscillation is caused by perturbations in the angular velocity of NS due to small disturbances in its density. The angular velocity $\omega$ of these r-mode oscillations, in the lowest order terms of an expansion in terms of $\Omega$, is given by \cite{Andersson1999,Papaloizou1978,Lindblom1998}

\begin{equation}
\omega=-\frac{\left(l-1\right)\left(l+2\right)}{l+1} \Omega
\label{eq3}
\end{equation}
where $l$ defines the mode and $\Omega$ represents the unperturbed angular frequency of the NS in its inertial frame. In the present work, $l=2$ quadrupole r-mode which is most important has been studied.
		
    As the viscosity \cite{Lindblom1987} opposes the GW emission, the instability in the mode grows. To ensure the instability to be pertinent, it should increase faster than the viscous dampening. Thus the time scale for gravitation driven instability should be sufficiently short compared to time scale of viscous damping. The time dependence of r-mode amplitude evolves as $ e^{i\omega t - t/\tau }$ due to the joint influence of ordinary hydrodynamics and various dissipative processes \cite{Lindblom1998,Lindblom2000,Owen1998}. The associated time-scales of different processes involve the actual physical parameters such as mass, radius, core radius etc., of NS. While computing these physical parameters, nuclear physics plays a significant role in limiting the uncertainties in nuclear EoSs. 
 
    The balancing effect between the gravitational radiation and the dissipative influence of viscosity results in the evolution of the r-modes. The present work studies these effects on the energy evolution of the modes. The r-mode dissipative time scale $\tau$ can be expressed as \cite{Lindblom1998} 
\begin{equation}
\frac{1}{\tau(\Omega,T)}=\frac{1}{\tau_{GR}(\Omega,T)}+\frac{1}{\tau_{VE}(\Omega,T)}+\frac{1}{\tau_{SV}(\Omega,T)}+\frac{1}{\tau_{BV}(\Omega,T)},
\label{eq7}
\end{equation}
where $1/\tau_{GR}$, $1/\tau_{VE}$, $1/\tau_{SV}$ and  $1/\tau_{BV}$ are the contributions from gravitational radiation, viscous dissipation in the crust-core boundary layer, shear and bulk viscous time scales in the fluid core, respectively, and are given by \cite{Lindblom2000,Owen1998}
\begin{figure}[t]
\vspace{0.0cm}
\includegraphics[width=8cm]{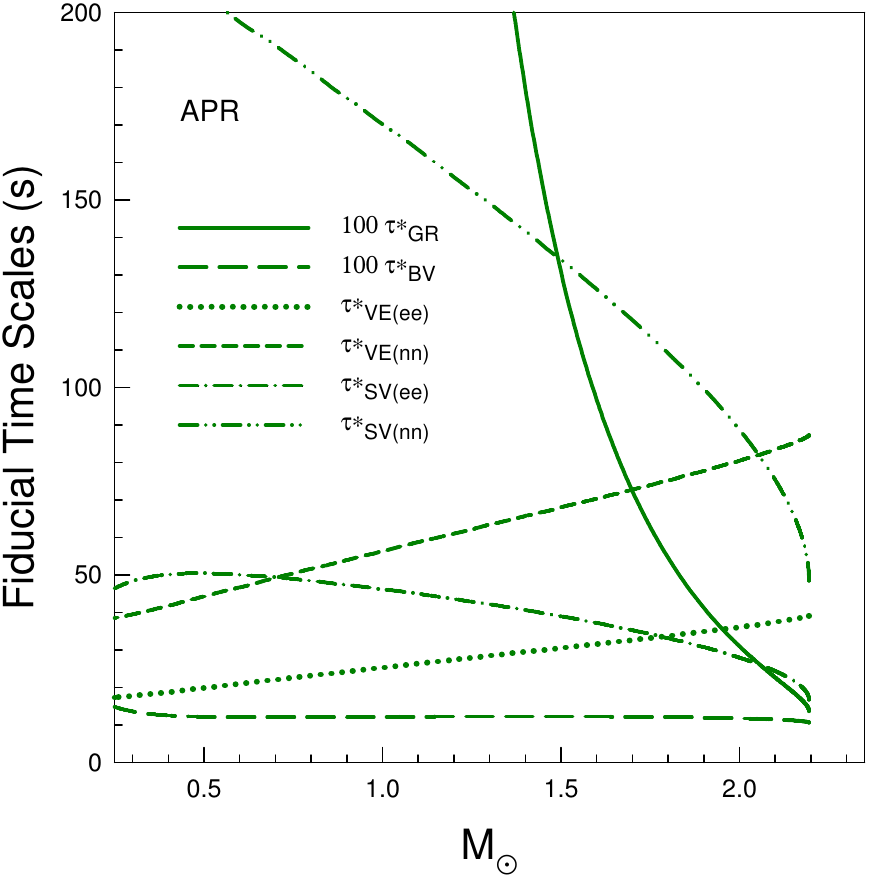}
\caption{The fiducial timescales plotted as a function of neutron star mass obtained using APR EoS.} 
\label{fig2}
\vspace{-0.0cm}
\end{figure} 

\begin{eqnarray}
\frac{1}{\tau_{GR}}=-\frac{32 \pi G \Omega^{2l+2}}{c^{2l+3}} \frac{(l-1)^{2l}}{[(2l+1)!!]^2}\left(\frac{l+2}{l+1}\right)^{(2l+2)}\nonumber \\
 \times\int^{R_{c}}_{0}\varrho(r)r^{2l+2} dr, 
\label{eq8}
\end{eqnarray}
for dissipation due to gravitational radiation where $\varrho(r)$ is the radial dependence of the mass density of NS. For the viscous dissipation in the crust-core boundary layer
\begin{eqnarray}
\frac{1}{\tau_{VE}}=\left[\frac{1}{2\Omega} \frac{2^{l+3/2}(l+1)!}{l(2l+1)!!I_{l}}\sqrt{\frac{2\Omega R_{c}^{2} \varrho_{c}}{\eta_c}}\right]^{-1}\nonumber \\
\times\left[\int^{R_{c}}_{0} \frac{\varrho(r)}{\varrho_{c}}\left(\frac{r}{R_{c}}\right)^{2l+2} \frac{dr}{R_c}\right]^{-1}. 
\label{eq9}
\end{eqnarray}
Here $\varrho_{c}$ and $\eta_{c}$ are the mass density and the shear viscosity of the fluid at the outer edge of the core having radius $R_{c}$. The shear viscous dissipation time scale in the fluid core is given by \cite{Lindblom1998}

\begin{equation}
\frac{1}{\tau_{SV}}=(l-1) (2l+1) \left(\int^{R_c}_{0}\varrho(r)r^{2l+2} dr\right)^{-1}\int^{R_c}_{0}\eta r^{2l} dr 
\label{eq10}
\end{equation}
where $\eta$ is the shear viscosity whereas for the bulk viscous time scales in the fluid core is given by

\begin{eqnarray}
\frac{1}{\tau_{BV}}&=& \frac{4\pi R^{2l-2}}{690} \Big(\frac{\Omega}{\Omega_0}\Big)^4 \left(\int^{R_{c}}_{0}\varrho(r)r^{2l+2} dr\right)^{-1} 
\nonumber\\
&\times&\int^{R_{c}}_{0}\xi_{BV}\Big(\frac{r}{R}\Big)^6 \Big[1+0.86\Big(\frac{r}{R}\Big)^2\Big] r^2dr 
\label{eq11}
\end{eqnarray}
where $\xi_{BV}$ is the bulk viscosity.

\begin{figure}[t]
\vspace{0.0cm}
\includegraphics[width=8cm]{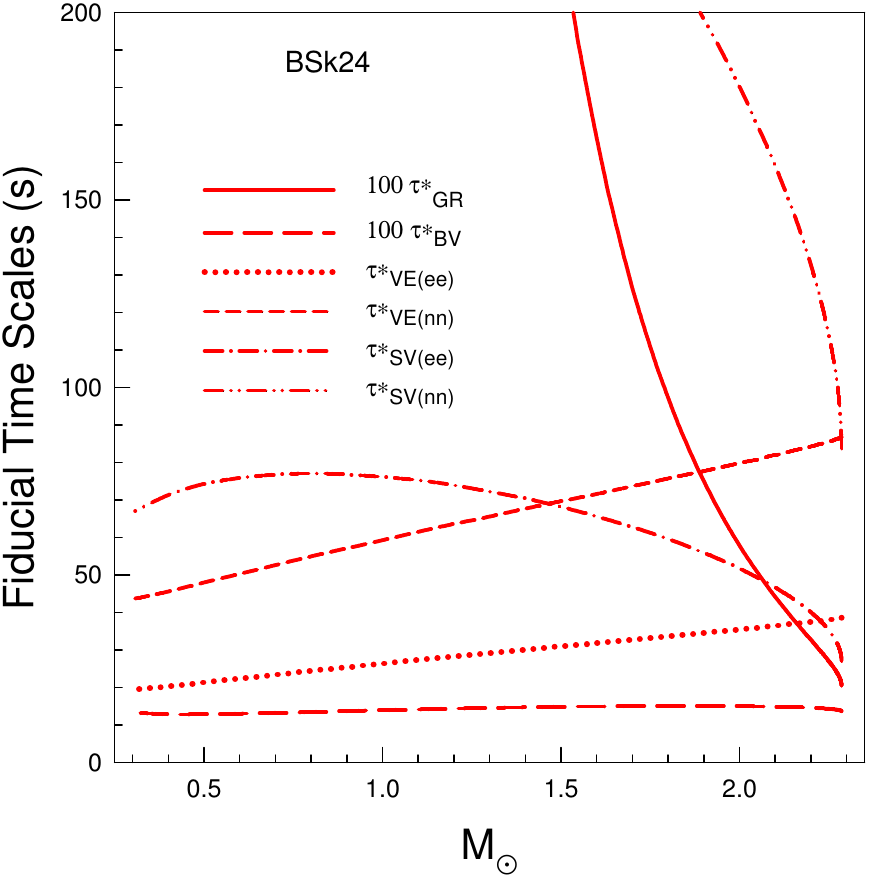}
\caption{The fiducial timescales plotted as a function of neutron star mass obtained using BSk24 EoS.} 
\label{fig3}
\vspace{-0.0cm}
\end{figure} 

    The time scale of shear viscosity described in Eq.~(\ref{eq9}) is determined by examining the dissipation due to shear viscosity within the boundary layer separating the solid crust and liquid core. This calculation assumes that the crust is inflexible and hence stationary within a rotating frame \cite{Lindblom2000}.
		
		The crustal motion resulting from its coupling with the core causes $\tau_{VE}$ to increase by a factor of $(\Delta v/v)^{-2}$ which is obtained by dividing the difference between the velocities of the outer edge of the core and the inner edge of the crust by the core's velocity \cite{Levin2001}.

    The impact of a solid crust on r-mode instability was initially calculated by Bildsten and Ushomirsky \cite{Bildsten2000}. They found that the shear dissipation within the viscous boundary layer resulted in reduction of the viscous damping time scale exceeding $10^5$ in accreting old NSs and surpassing $10^7$ in young and hot rotating NSs. At higher temperatures the shear viscosity contribution is dominated by neutron-neutron (nn) scattering, whereas at lower temperatures the electron-electron (ee) scattering is dominant \cite{Lindblom2000}. Thus    		

\begin{equation}
\frac{1}{\tau_{VE(SV)}}=\frac{1}{\tau_{ee}}+\frac{1}{\tau_{nn}},
\label{eq12}
\end{equation}
where $\tau_{ee}$ and $\tau_{nn}$ are evaluated from Eqs.~(\ref{eq9},\ref{eq10}) using the corresponding values of $\eta_{ee}$ and $\eta_{nn}$ given in \cite{Flowers1979,Cutler1987,Sawyer1989,Moustakidis2016}:
\begin{equation}
\left(\frac{\eta_{ee}}{\rm g~cm^{-1}s^{-1}}\right)=\mathcal{A}_{ee} \left(\frac{\varrho}{\rm  g~cm^{-3}}\right)^{2} \left(\frac{T}{K}\right)^{-2}, 
\label{eq13}
\end{equation}

\begin{equation}
\left(\frac{\eta_{nn}}{\rm g~cm^{-1}s^{-1}}\right)=\mathcal{A}_{nn} \left(\frac{\varrho}{\rm  g~cm^{-3}}\right)^{9/4} \left(\frac{T}{K}\right)^{-2}, 
\label{eq14}
\end{equation}
 
\begin{equation}
\left(\frac{\xi_{BV}}{\rm g~cm^{-1}s^{-1}}\right)=\mathcal{A}_{BV} \left(\frac{l+1}{2}\right)^{2} \left(\frac{\rm Hz}{\Omega}\right)^{2} \left(\frac{\varrho}{\rm  g~cm^{-3}}\right)^{2}\left(\frac{T}{K}\right)^{6}
\label{eq15}
\end{equation} 
where the constants $\mathcal{A}_{ee}= 6 \times 10^{6}$, $\mathcal{A}_{nn}= 347$ and $\mathcal{A}_{BV}= 6 \times 10^{-59}$. 

\begin{figure}[t]
\vspace{0.0cm}
\includegraphics[width=8cm]{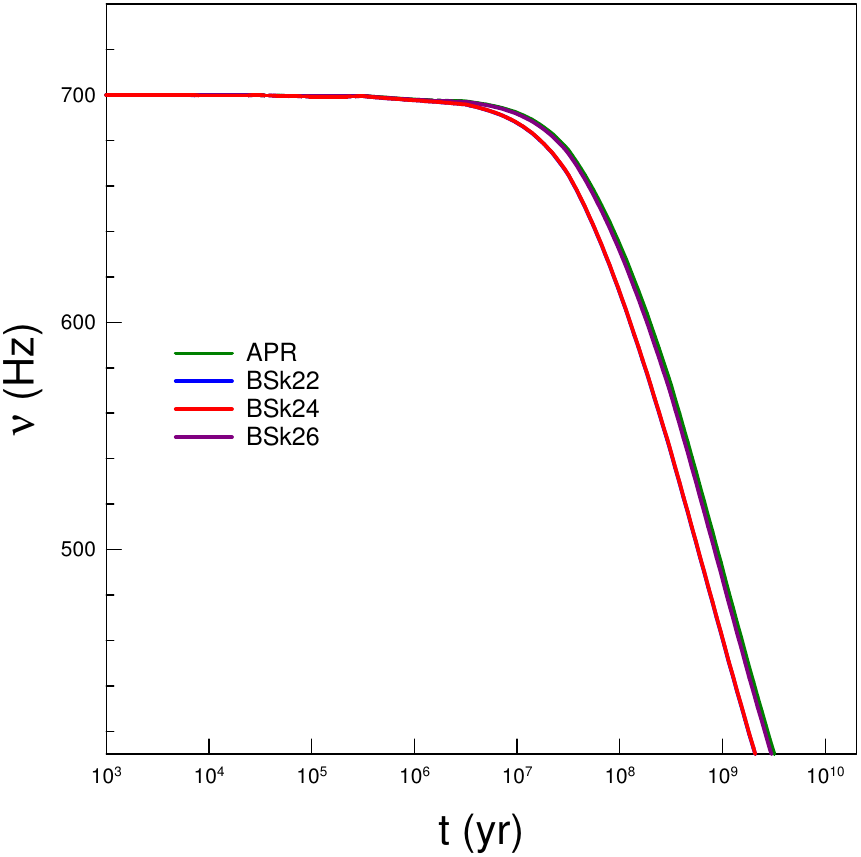}
\caption{Plots of frequencies as a function of time.} 
\label{fig4}
\vspace{-0.0cm}
\end{figure}

    The dissipation effects due to viscosity cause the $r$-mode to decay exponentially as $e^{-t/\tau}$ as long as $\tau > 0$ \cite{Lindblom1998}. In order to facilitate transparent visualization of the roles of temperature $T$ and the angular velocity $\Omega$ in various time-scales, it is convenient to separate them by defining their respective fiducial time-scales. The time-scale $\tau$ given in the Eq.~(\ref{eq7}) can be expressed as       

\begin{eqnarray}
\frac{1}{\tau(\Omega,T)}=\frac{1}{\tau^*_{GR}}\left(\frac{\Omega}{\Omega_0}\right)^{2l+2}+\frac{1}{\tau^*_{VE}}  \left(\frac{10^8 K}{T}\right) \left(\frac{\Omega}{\Omega_0}\right)^{1/2}
\nonumber \\
+\frac{1}{\tau^*_{SV}}\left(\frac{10^6 K}{T}\right)^{2}
+\frac{1}{\tau^*_{BV}}\left(\frac{\Omega}{\Omega_0}\right)^{2}\left(\frac{T}{10^{11} K}\right)^{6}
\label{eq16}
\end{eqnarray}
where $\Omega_0=\sqrt{ \pi G \bar{\rho}}$, with $\bar{\rho}= 3M/4 \pi R^3$ being the mean density of a NS with mass $M$ and radius $R$. ${\tau^*_{GR}}$, ${\tau^*_{SV}}$, ${\tau^*_{BV}}$ and ${\tau^*_{VE}} $ are the respective fiducial time-scales that can be defined from Eq.~(\ref{eq7}) and Eq.~(\ref{eq16}). It is evident from Eqs.~(\ref{eq8}-\ref{eq11}) that $\tau^*_{VE}$, $\tau^*_{SV}$, $\tau^*_{BV}$ are positive while $\tau^*_{GR}$ is negative. As a result viscosity has a stabilizing effect while gravitational radiation pushes these modes towards instability. Since the gravitational radiation is proportional to $\Omega^{2l+2}$, its impact on $1/\tau$ is very low for small $\Omega$. It implies that viscosity will dominate and the mode will be stable for sufficiently small angular velocities. On the contrary, for large values of $\Omega$ the gravitational radiation dominates and drives the mode towards instability. For a particular mode $l$ and a given temperature $T$, the condition $1/\tau(\Omega_c,T)=0$ defines the critical angular velocity $\Omega_c$. 
    
\begin{figure}[t]
\vspace{0.0cm}
\includegraphics[width=8cm]{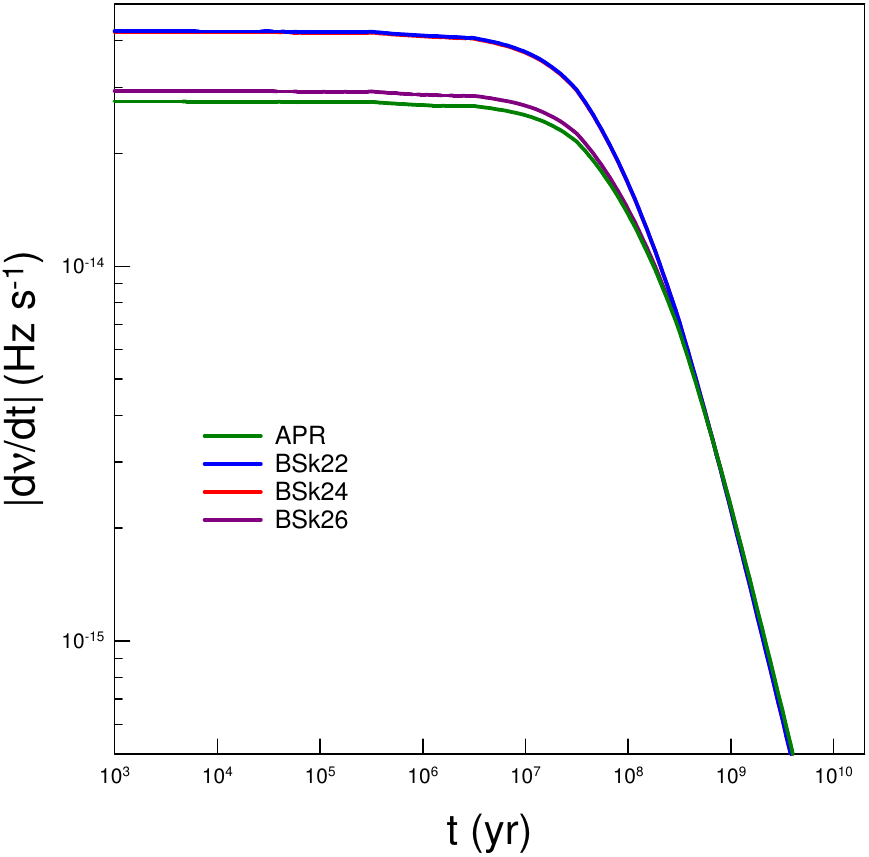}
\caption{Plots of spin-down rates as a function of time.} 
\label{fig5}
\vspace{-0.0cm}
\end{figure} 
		
    Since that the angular momentum is radiated to infinity by the gravitational radiation, the angular velocity evolves with time as \cite{Owen1998} 
    
\begin{equation}
\frac{d\Omega}{dt}=\frac{2\Omega}{\tau_{GR}}\frac{\alpha_r^2Q}{1-\alpha_r^2Q}.
\label{eq17}
\end{equation}
The dimensionless quantities $\alpha_r$ is the r-mode amplitude and $Q=3 \widetilde{J}/2 \widetilde{I}$ where
\begin{equation}
\widetilde{J}=\frac{1}{MR^4}\int^{R}_{0}\varrho(r)r^{6} dr
\label{eq18}
\end{equation}
and
\begin{equation}
\widetilde{I}=\frac{8\pi}{3MR^2}\int^{R}_{0}\varrho(r)r^{4} dr.
\label{eq19}
\end{equation}
The values of $\alpha_r$ can be obtained from thermal equilibrium or spin equilibrium or may be treated as a free parameter with values ranging from $1$ to $10^{-8}$. The solution for the angular frequency $\Omega(t)$ can be obtained from Eq.~(\ref{eq17}) as

\begin{equation}
\Omega(t)=\left(\Omega^{-6}_{in}-\mathcal{C}t\right)^{-1/6},
\label{eq20}
\end{equation}
by considering that the heat extracted from neutrino emission is same \cite{Bondarescu2009,Moustakidis2015} as that generated by shear viscosity under ideal conditions, where 

\begin{equation}
\mathcal{C}=\frac{12\alpha_r^2Q}{\tau^*_{GR}\left(1-\alpha_r^2Q\right)}\frac{1}{\Omega_0^6}.
\label{eq21}
\end{equation}

The quantity $\Omega_{in}$, whose value conforms to the initial angular velocity, has been treated as a free parameter. Using Eqs.~(\ref{eq20},\ref{eq21}), the spin down rate given by Eq.~(\ref{eq17}) reduces to

\begin{equation}
\frac{d\Omega}{dt}=\frac{\mathcal{C}}{6}\left(\Omega^{-6}_{in}-\mathcal{C}t\right)^{-7/6}.
\label{eq22}
\end{equation}

    The spin of a NS decreases continuously till it reaches $\Omega_c$. The time $t_c$ elapsed for a NS evolving from the initial angular velocity $\Omega_{in}$ to $\Omega_{c}$ which is its minimum value is given by
    
\begin{equation}
t_c=\frac{1}{\mathcal{C}}\left(\Omega_{in}^{-6}-\Omega_{c}^{-6}\right).
\label{eq23}
\end{equation}		

\begin{figure}[t]
\vspace{0.0cm}
\includegraphics[width=8cm]{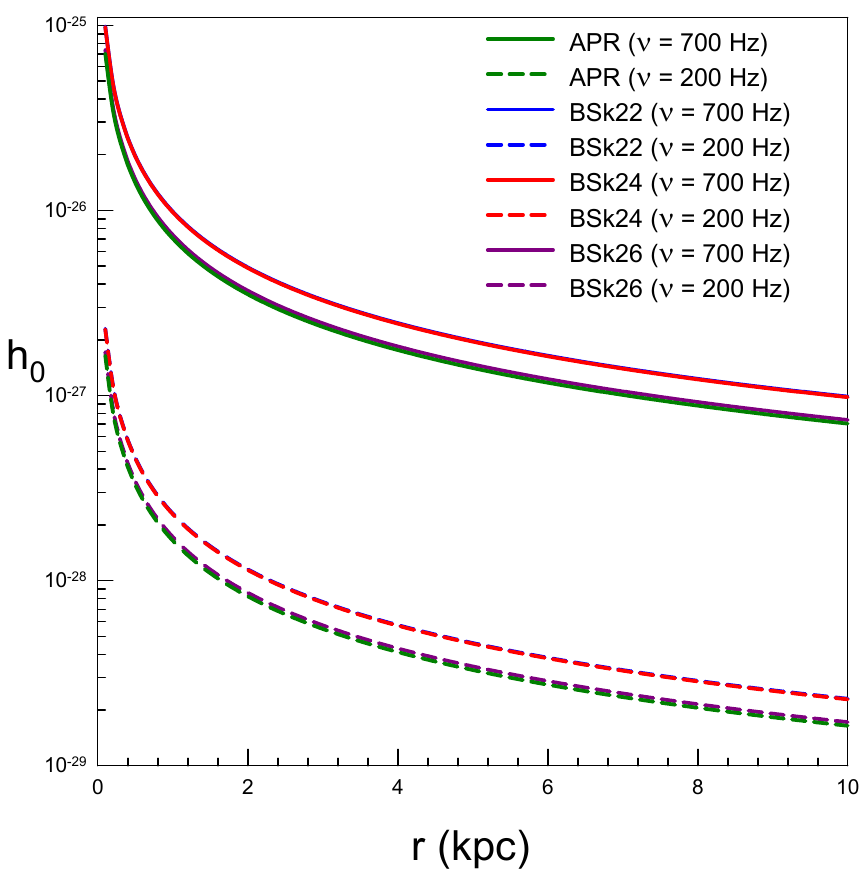}
\caption{Plots displaying the amplitude of the strain tensor ($h_0$) in relation to the distance for two distinct rotational frequencies of neutron stars, using r-mode amplitude as $\alpha_r=2 \times 10^{-7}$ .} 
\label{fig6}
\vspace{-0.0cm}
\end{figure} 

    As mentioned earlier, the r-mode instability arises when the time scale associated with gravitational-radiation is shorter than the dissipation time scales of various mechanisms present within a NS's interior. The nuclear EoS has two distinct impacts on the r-mode time scales. Firstly, it defines the radial distribution of mass density, which is the fundamental ingredient in the integrals that are relevant to the r-mode. Secondly, the EoS specifies the transition density from the NS core to its crust and the radius of the core, which act as the upper limits of these integrals. 

    In this work EoSs obtained using APR, BSk22, BSk24 and BSk26 interactions have been used. The $\beta$-equilibrated $npe\mu$ neutron star matter has been used for subsequent calculations. The observations of the binary millisecond pulsar J1614-2230 suggest that its mass lies in the range $1.97\pm0.04$ M$_\odot$ \cite{De10}. The measurements of radio timing for pulsar PSR J0348+0432 and its companion (white dwarf) have confirmed the mass of the pulsar to be in the range of 1.97$-$2.18 $M_\odot$ \cite{An13}. Very recently, the studies for PSR J0740+6620 \cite{Fon21} and for PSR J0952-0607 \cite{Rom22} find masses of 2.08 $\pm$ 0.07 $M_\odot$ and 2.35 $\pm$ 0.17 $M_\odot$, respectively. Current observations of PSR J0740+6620 also suggest its mass to be 2.072$^{+0.067}_{-0.066}$ $M_\odot$ \cite{Leg21,Ril21}. The maximum masses obtained using APR, BSk22, BSk24 and BSk26 are respectively, 2.19 $M_\odot$, 2.27 $M_\odot$, 2.28 $M_\odot$ and 2.18 $M_\odot$ \cite{La24} which provide reasonably good estimates for latest observations of NS masses.  
  
\section{Gravitational waves from rotating compact stars}
         
\subsection{Constraints from Thermal Equilibrium}

    The gravitational radiation, in a steady-state, injects energy into the r-mode at a rate given by \cite{Mahmoodifar2013}
    
\begin{equation}
 W_d =  (1/3)\Omega \dot J_c = -2\widetilde{E}/\tau_{GR}.
\label{eq24}
\end{equation} 
\noindent
This energy, in the condition of thermal steady-state, dissipates totally within the star. The energy $L_\nu$ is lost due to the neutrino emission while the rest due to emission of photons $L_\gamma$ \cite{Bondarescu2009,Moustakidis2015}, both of which get radiated from the star's surface. It is important to mention that when the mode becomes saturated \cite{Alford2014}, the thermal steady-state is independent of the cooling mechanism. Thus the thermal steady-state condition is $W_d=L_\nu+L_\gamma$ assuming that during quiescence all the emitted energy is entirely due to dissipation of r-mode within the star. The $J_c$ appearing in the Eq.~(\ref{eq24}) is the canonical angular momentum of the mode given by $J_c = -(3/2)\widetilde{J}M R^2 \Omega \alpha_r^2$, where $\widetilde{J}$ and $\widetilde{I}$ are the dimensionless quantities defined in Eqs.~(\ref{eq18},\ref{eq19}), respectively \cite{Mu18}. Using Eq.~(\ref{eq8}) and the explicit expression for $J_c$, the amplitude at saturation $\alpha_{th}$ in thermal equilibrium is then given by

\begin{equation}
\alpha_{th} =\Big[\frac{-\tau_{GR}4\pi R^2\sigma T_{eff}^4}{\widetilde{J}M}\Big]^{1/2}\frac{1}{\Omega R},
\label{eq25}
\end{equation} 
\noindent
where $\sigma$ is the Stefan's constant. 

\begin{figure}[t]
\vspace{0.0cm}
\includegraphics[width=8cm]{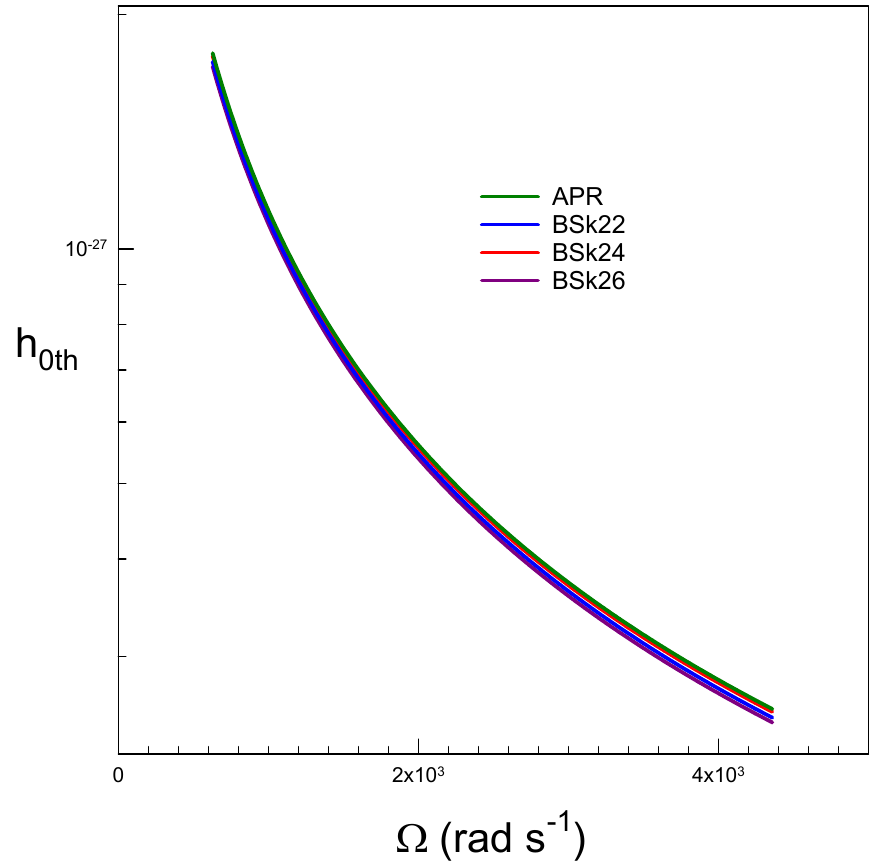}
\caption{Plots displaying the amplitude of the strain tensor ($h_{0{\rm th}}$) in relation to the angular frequency of the neutron stars, while the r-mode amplitude estimated from the thermal equilibrium.} 
\label{fig7}
\vspace{0.0cm}
\end{figure} 
				
\begin{figure}[t]
\vspace{0.0cm}
\includegraphics[width=8cm]{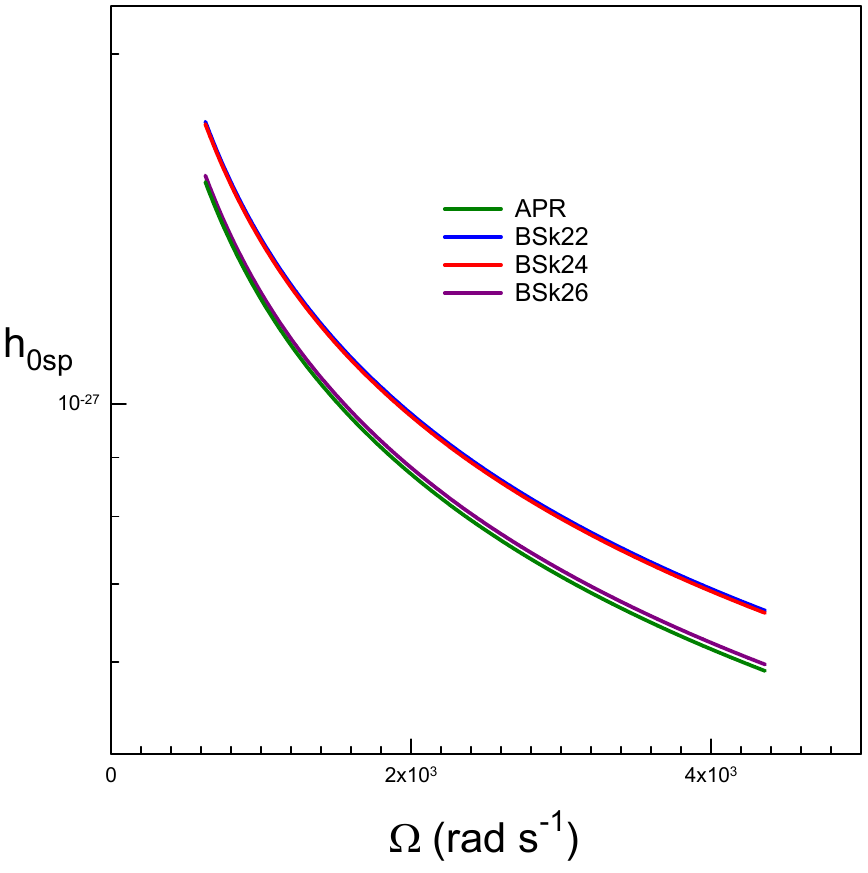}
\caption{Plots displaying the amplitude of the strain tensor ($h_{0{\rm sp}}$) in relation to the angular frequency of the neutron stars, while the r-mode amplitude estimated from spin equilibrium.} 
\label{fig8}
\vspace{-0.0cm}
\end{figure}        		
         		
\subsection{Constraints from Spin Equilibrium}
    
    The NSs are sources of extremely high gravitational attraction towards the center due to their very high compactness. As a consequence, these stars can acquire mass from the companion if they belong to a binary system. Large angular momentum gets absorbed since any mass moving around the star can be free only at angular velocities reaching the Keplerian limit. The NS gains angular momentum by absorbing such masses of the companion star which is rotating with speed much lower than the Keplerian angular velocity. Thus a NS gradually spins up as it accumulates mass from its binary companion at a rate of $\dot M$. Their ejecta and observed spin-up rates may be used to constrain the amplitude of r-modes \cite{Brown2000, Ho2011}. Therefore
    
\begin{equation}
2\pi I \dot{\nu} \Delta =  \frac{2J_c}{\tau_{GR}}
\label{eq26},
\end{equation} 
\noindent
where $I=MR^2\widetilde{I}$ is the moment of inertia. This equation displays the effect of r-mode perturbation on the right hand side, specifically the torque resulting from gravitational emission causing spin-down. The quantity $\dot{\nu}$ is the spin-up rate during outburst and $\Delta = (t_o/t_r)$ is the ratio of the outburst duration $t_o$ to the recurrence time $t_r$. Using Eq.~(\ref{eq26}) and the explicit expression for $J_c$, the amplitude at saturation $\alpha_{sp}$ in spin equilibrium follows:  
		
\begin{equation}
\alpha_{sp} =\Big[\frac{-\tau_{GR}2\pi \widetilde{I} \dot{\nu}\Delta}{3\widetilde{J}\Omega}\Big]^{1/2}
\label{eq27}
\end{equation} 
\noindent

\noindent     
\subsection{Gravitational Wave Amplitudes} 
 
     The angular momentum gets transferred as a NS accretes mass from its companion causing increase in its rate of rotation. Ultimately it surpasses its critical value $\Omega_c$. The NS begins emitting GW at this epoch due to r-mode perturbation. The GW emission fuels this perturbation due to r-mode because of Chandrasekhar–Friedman–Schutz (CFS) mechanism \cite{Chandrasekhar1970}, resulting in an increase in the amplitude $\alpha_r$ until it saturates. As described earlier, its saturation value may be estimated either from thermal equilibrium or spin equilibrium. The NS spins down to the region of stability by emitting GWs which take away the energy and the angular momentum with it. In the present calculations the emitted GW intensity radiated by NSs is described in terms of the strain tensor amplitude $h_0$, which in turn is related to the amplitude $\alpha_r$ of the $r$-mode as \cite{Owen2009, Owen2010}  

\begin{eqnarray}
h_0=\sqrt{\frac{8\pi}{5}} \frac{G}{c^5} \frac{1}{r} \alpha_r \omega^3 M R^3\widetilde{J}.
\label{eq28}
\end{eqnarray} 
\noindent
It is worthwhile to mention that the signature of continuous emission of GW because of the perturbation due to r-mode is different from GW emission due to the ellipticity of a star \cite{Owen2010}. In case of r-mode perturbation, the GW emission is dominated by the mass quadrupole moment, while for ellipticity, it is through the mass current quadrupolar moment. 

\begin{figure}[t]
\vspace{0.0cm}
\includegraphics[width=8cm]{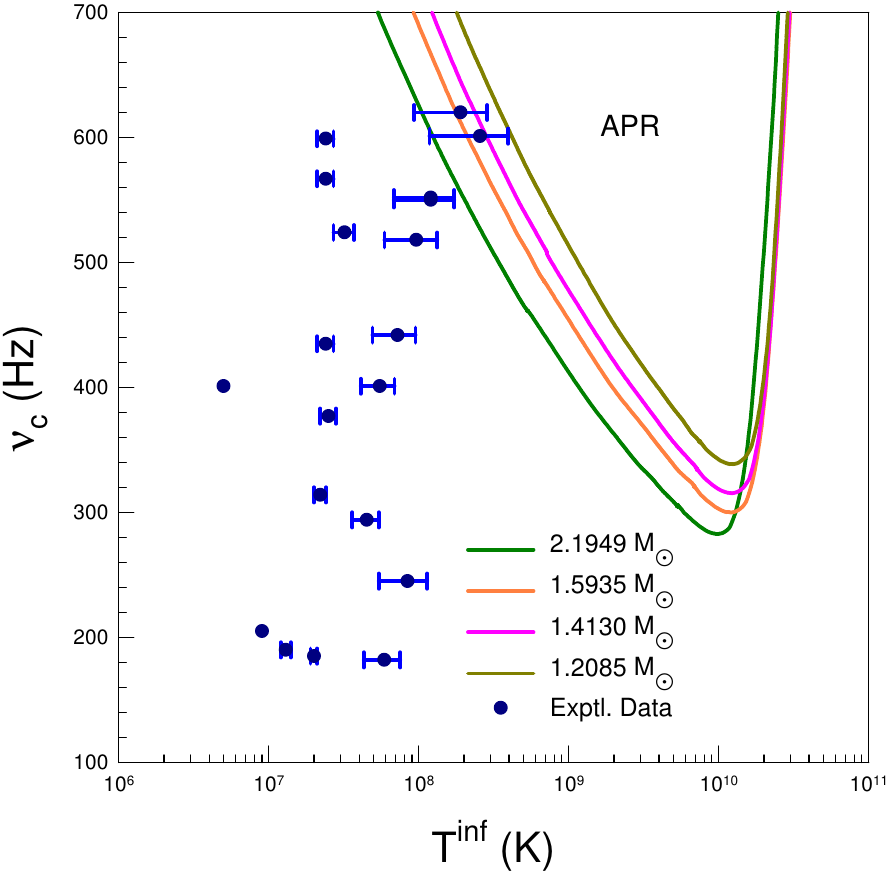}
\caption{The critical spin frequency plotted as a function of temperature for neutron stars of different gravitational masses obtained using APR EoS. The observational data \cite{Gus14} are represented by dots along with error bars.} 
\label{fig9}
\vspace{-0.0cm}
\end{figure} 
    
\noindent
\section{Results and discussion}
\label{Section 6}

    In this work, EoSs obtained from the effective nucleon-nucleon (NN) interaction described by APR \cite{Ak98} and Brussels-Montreal Skyrme effective interaction with the BSk22, BSk24 and BSk26 parameter sets \cite{Gor13} have been used. The $\beta$-equilibrated $npe\mu$ neutron star matter has been used for subsequent calculations. The calculations for masses and radii have been performed using Feynman-Metropolis-Teller   
[FMT] \cite{FMT}, Baym-Pethick-Sutherland [BPS] \cite{Ba71} and Baym-Bethe-Pethick [BBP] \cite{Ba71a} the EoSs up to the number density of 0.0582 fm$^{-3}$ for the outer crust and the $\beta$-equilibrated NS matter beyond which takes care of the inner crustal and core regions of a compact star. The location of the inner edge of the NS crust, the core-crust transition density and pressure were determined \cite{La24} by dynamical method. The mass versus radius of NSs obtained using APR, BSk22, BSk24 and BSk26 EoSs have been plotted in Fig.-\ref{fig1}. The maximum masses obtained using APR, BSk22, BSk24 and BSk26 EoSs are 2.19 $M_\odot$, 2.27 $M_\odot$, 2.28 $M_\odot$ and 2.18 $M_\odot$, respectively. The shaded regions represent the HESS J1731-347 remnant \cite{Dor22}, the GW170817 event \cite{Abbott19}, PSR J1614-2230 \cite{Arz18}, PSR J0348+0432 \cite{Ant13}, PSR J0740+6620 \cite{Cro20}, and PSR J0952-0607 \cite{Rom22} pulsar observations for the possible maximum mass. The black dots with error bars represent recent radius measurements from the Neutron Star Interior Composition Explorer (NICER) \cite{Ril21,Ril19}. The fiducial timescales versus the gravitational masses of NSs have been plotted in Fig.-\ref{fig2} and Fig.-\ref{fig3} for the APR and BSk24 EoSs, respectively. While the timescale for gravitational radiation falls off rapidly, it is observed that the timescales for viscous damping $\tau^*_{VE}$ at the core-crust interface increase almost linearly with increasing mass. While the effect of bulk viscosity $\tau^*_{BV}$ is almost independent of NS mass, the timescales for viscous damping $\tau^*_{SV}$ in the core also decreases but at a somewhat slower rate than that for gravitational radiation. 

\begin{figure}[t]
\vspace{0.0cm}
\includegraphics[width=8cm]{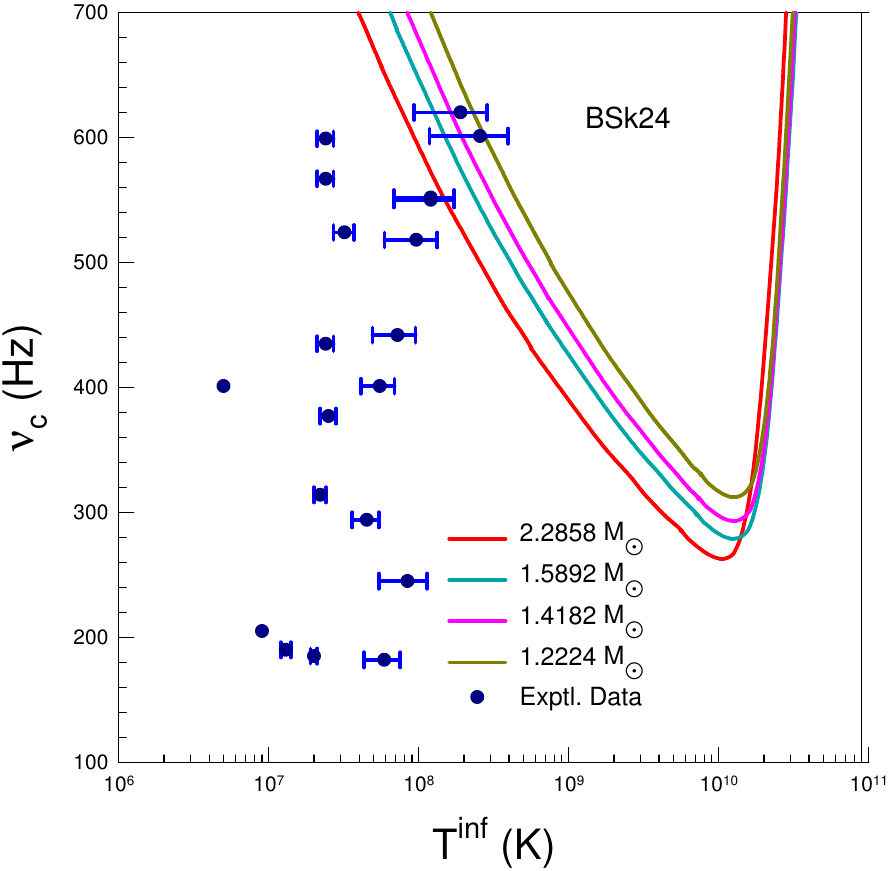}
\caption{The critical spin frequency plotted as a function of temperature for neutron stars of different gravitational masses obtained using BSk24 EoS. The observational data \cite{Gus14} are represented by dots along with error bars.} 
\label{fig10}
\vspace{-0.0cm}
\end{figure}
 		
		The dependence of the critical angular velocity $\Omega_c$ on temperature for $(l=2)$ r-mode may be explored by using the fiducial timescales of gravitational radiation and shear viscous drag. 
		
    The amplitude $\alpha_r$ of the r-mode grows till it attains a saturation value when a NS enters into the region of instability because of mass accretion from its binary companion. At this point a NS emits GWs releasing its energy and angular momentum leading to spin down to reach the stability region. The time evolution of angular velocity of spinning and the rate of spin down can be calculated for a NS from Eq.~(\ref{eq20}) and Eq.~(\ref{eq22}), respectively, assuming the ideal condition that the decrease in temperature due to emission of GWs is compensated by the heat produced due to viscous effects, provided T, M, $\alpha_r$ and $\Omega_{in}$ of the star are known. The spin evolution is calculated corresponding to maximum NS masses as obtained from APR, BSk22, BSk24 and BSk26 EoSs, using the representative values of $\nu_{in}=\Omega_{in}/(2\pi)=700$ Hz and $\alpha_r=2 \times 10^{-7}$ as used by Moustakidis \cite{Moustakidis2015} and shown in Fig.-\ref{fig4}. The rates of spin down are shown for these masses in Fig.-\ref{fig5}. 
		
    The plots of strain tensor amplitude $h_0$ as a function of distance for four different maximum NS masses for two spin frequencies of 200 Hz and 700 Hz are shown in Fig.-\ref{fig6}, using r-mode amplitude as $\alpha_r=2 \times 10^{-7}$. The quantity $\alpha_r$ is estimated from thermal equilibrium as well using the red-shifted effective surface temperature $T^{\rm inf} = 100$ eV. For the same $T^{\rm inf}$, effective surface temperature for higher mass would be higher which sets the trend resulting from higher $\alpha_r$ and correspondingly higher strain tensor amplitude $h_{\rm 0th}$. In Fig.-\ref{fig7}, the strain tensor amplitude $h_{\rm 0th}$ is plotted as a function of evolving angular velocity $\Omega$ of the star in units of rad s$^{-1}$ for four different maximum NS masses which arise from GW emissions. In Fig.-\ref{fig8} the strain tensor amplitude $h_{\rm 0sp}$ is plotted as a function of angular velocity $\Omega$ with $\alpha_r$ estimated from spin equilibrium using Eq.~(\ref{eq27}) where the experimental data used were from the source IGR J00291 \cite{Mahmoodifar2013}. Entire calculations have been performed presuming the range of angular momentum of the star such that the CFS two-stream instability \cite{Chandrasekhar1970} applies enhancing the r-mode perturbation as the GW emission pumps into the perturbation instead of damping it. Thus the NS, spinning at a particular angular velocity emits GW because of the perturbation and causing loss of its angular momentum. As a consequence while the the star's angular velocity continues to reduce till it crosses the critical limit ($\Omega_c$) of CFS two-stream instability, the intensity of GW emission increases.  

    The frequencies of spinning and the temperatures of the core (either measured or the estimated upper limits) of the observed Low Mass X-ray Binaries (LMXBs) and Millisecond Radio Pulsars (MSRPs) \cite{Gus14} have been shown in Fig.-\ref{fig9} and Fig.-\ref{fig10} along with critical frequency versus temperature plots for the APR and BSk24 EoSs, respectively. The Figs.-\ref{fig9},\ref{fig10} imply that all of the observed NSs lie in the stable region of the r-mode oscillations defined using APR, BSk22, BSk24 and BSk26 EoSs with a rigid crust and have comparatively small amplitudes of r-mode oscillations, which is consistent with the fact that GW emission from these NSs due to r-mode instability have not been observed.	For a fixed temperature $T$, $\Omega_c/\Omega_0$ rapidly decreases with increasing mass. 	
       		
\noindent
\section{Summary and conclusions}
\label{Section 7}

    In the present work, the Rossby mode oscillations in pulsars, their role in continuous GW emission and the effects of viscous damping have been explored. The fiducial gravitational radiation timescale and the timescales for the bulk and shear viscous drags at the crust-core boundary have been calculated using $\beta$-equilibrated nuclear matter EoSs obtained in the framework of APR and Skyrme effective nucleon-nucleon interaction with the BSk22, BSk24 and BSk26 parametrization for a wide range of pulsar masses. Present study identifies critical angular velocities where r-modes become unstable, highlighting how gravitational radiation drives oscillations toward instability while viscous damping stabilizes them at low angular velocities. The r-mode oscillations evolve on a time-scale much shorter than that associated with the damping due to viscosity. The viscous boundary layer between the oscillating fluid in the core and the co-rotating crust shows that energy dissipation is the predominant r-mode damping mechanism. Using these EoSs, it is observed that while the shear viscous damping timescales  $\tau_{VE}$ increase almost linearly with increasing NS mass, the GW emission timescales decrease rapidly. The pulsars spinning faster than their corresponding critical frequencies have unstable r-modes leading to the emission of GWs. The intrinsic strain tensor amplitude $h_0$ has been derived by assessing the amplitude $\alpha_r$ of the r-mode perturbation, considering both spin equilibrium and thermal equilibrium. Spin equilibrium is applicable to stars that are active and accreting matter from their binary partners. Conversely, the thermal equilibrium concept is more appropriate for stars that have ceased accretion or those that do not exist within a binary system. Thus, it is anticipated that the amplitude of the strain tensor obtained from spin equilibrium would be greater than that derived from thermal equilibrium.

    The r-mode instability is of interest for rotating neutron stars as it is of astrophysical relevance due to the ongoing current experiments for detecting GWs that may be detected by ground based interferometric detectors. The frequency range utilized in this work falls within the spectrum that the LIGO-Virgo collaboration has targeted for their searches for continuous GWs. A comparison with LIGO-Virgo data clearly indicates that the computed values of $h_0$ in this study exceed the detection limits of advanced LIGO. The detectors at the LIGO-Virgo facilities are undergoing continuous upgrades and enhancements. These ongoing improvements enhance the sensitivity of detectors to smaller gravitational ripples and their ability to minimize noise, thereby increasing the potential to detect lower values of intrinsic strain. This advancement may facilitate the identification of sources exhibiting r-mode perturbations in future.  
	
\vspace{0.5cm}    
\noindent
{\bf Acknowledgements}
\vspace{0.5cm}

    One of the authors (DNB) acknowledges support from Anusandhan National Research Foundation (erstwhile Science and Engineering Research Board), Department of Science and Technology, Government of India, through Grant No.CRG/2021/007333.



\end{document}